\documentclass[12pt]{iopart}
\usepackage{graphicx}

\begin{document}
\title[Current fluctuations in the TASEP]{Finite size scaling of current fluctuations in the totally asymmetric exclusion process}

\author{Mieke Gorissen$^{1}$ and Carlo Vanderzande$^{1,2}$}

\address{$^1$Faculty of Sciences, Hasselt University, 3590 Diepenbeek, Belgium}
\address{$^2$Instituut Theoretische Fysica, Katholieke Universiteit Leuven, 3001 Heverlee, Belgium}
\ead{mieke.gorissen@uhasselt.be, carlo.vanderzande@uhasselt.be}
\begin{abstract}
We study the fluctuations of the current $J(t)$ of the totally asymmetric exclusion process with open boundaries. Using a density matrix renormalization group approach, we calculate the cumulant generating function of the current. This function can be interpreted as a free energy for an ensemble in which histories are weighted  by $\exp(-sJ(t))$. We show that in this ensemble the model has a first order space-time phase transition at $s=0$.
We numerically determine the finite size scaling of the cumulant generating function near this phase transition, both in the non-equilibrium steady state and for large times.\\


\end{abstract}

\maketitle
\section{Introduction}
Heat or particle currents arise in systems that are driven away from equilibrium by bringing them in contact with reservoirs at different temperatures or chemical potentials. In macroscopic systems, fluctuations of these currents can often be neglected and a description using non-equilibrium thermodynamics is then appropriate \cite{deGroot84}. However, current fluctuations can become important in mesoscopic systems and in the vicinity of non-equilibrium critical points.

In recent years, the interest in current fluctuations has increased considerably. From the experimental side, it is nowadays possible to measure third- and higher-order cumulants of the current in problems of charge transport \cite{Heikkila07,Fujisawa06}. In quantum point contacts, current fluctuations can be used as an entanglement meter \cite{Levitov09}. Theoretically, one is interested in the entropy produced in the presence of a current, since its fluctuations have been shown to obey various kinds of fluctuation theorems \cite{Evans93,Gallavotti95,Kurchan98,Lebowitz99,Maes99,Seifert05,Esposito10}. 

In this article, we study current fluctuations in the totally asymmetric exclusion process (TASEP). In this classical, stochastic model, particles on a  one-dimensional lattice jump to the right (provided that site is empty), thus giving rise to a time-dependent particle current. The TASEP was originally introduced to describe the biological process of translation, in which ribosomes move along a messenger RNA and use the information stored on it to build proteins \cite{MacDonald69}. In this context, one can relate current fluctuations to fluctuations in protein production as we recently showed \cite{Gorissen10}. Nowadays, the TASEP, and some of its variants, have become paradigmatic for non-equilibrium statistical mechanics \cite{Derrida07}. 

Fluctuations in the current can be determined from the cumulant generating function (CGF). In the large time limit, the TASEP evolves to a non-equilibrium steady state (NESS) for which the CGF equals the largest eigenvalue of a generalized generator \cite{Derrida07}. This largest eigenvalue has been determined exactly for the TASEP on a ring \cite{Derrida98,Derrida99} using the Bethe-ansatz and was found to have an interesting scaling form, in which the KPZ-dynamical exponent $z=3/2$ \cite{Kardar86,Halpin-Healy95} appears. Much less is known for a configuration with open boundaries where particles can enter on the left, and leave on the right. The average and variance of the current have been determined using a matrix approach \cite{Derrida93,Derrida95}, but no results are known for higher cumulants or for the full generating function. Recently, we showed how the latter  can be determined numerically using a density matrix renormalization group (DMRG) approach \cite{Gorissen09}. In this article, we further analyse our results and show that in the thermodynamic limit, the CGF has a non-analytic behaviour which corresponds to a space-time phase transition of a type that has been discovered recently in models of glassy dynamics \cite{Merolle05,Jack06,Garrahan07,Garrahan09}. We establish the scaling form of the CGF near this phase transition.

Using the same DMRG approach we also investigate the gap in the spectrum of the generalised generator. This gap determines the approach of the current fluctuations to their NESS-value. We investigate the scaling properties of the gap and conjecture a scaling form for the {\it time dependent current fluctuations} which so far have been hardly studied, but which could be relevant, for example, in applications to protein production \cite{Gorissen10}. The scaling form is verified with  simulations based on the Gillespie algorithm.

This paper is organized as follows. In section 2 we introduce the model, the cumulant generating function of the current, and the reinterpretion of this function as a free energy in the so called $s$-ensemble.
In section 3, we show that the CGF must be a non-analytical function of the variable $s$. In section 4 we discuss the DMRG approach. In the section 5 and 6 we present the results of our numerical calculations on the CGF and the time dependence of the current fluctuations respectively. We pay particular attention to the scaling behaviour of these quantities. Finally, in section 7 we present our conclusions.

\section{The cumulant generating function and the $s$-ensemble for the TASEP}
In the totally asymmetric exclusion process, each site of a one-dimensional lattice of $L$ sites can be empty or occupied by at most one particle. 
The dynamics of the TASEP is that of a continuous time Markov process for which the probability $P(C,t)$ that the system is in a configuration $C$ at time $t$ evolves according to the master equation
\begin{eqnarray}
\partial_t P(C,t) = \sum_{C'\neq C} \left[w(C'\to C) P(C',t)-w(C \to C') P(C,t)\right]
\label{1}
\end{eqnarray}
Here $w(C\to C')$ is the transition rate from configuration $C$ to $C'$. In the TASEP, particles can only  jump to the right (provided that site is empty) with unit rate. We impose open boundary conditions for which particles can enter the lattice on the left-side with rate $\alpha\leq1$, and leave it on the right with rate $\beta\leq 1$.  For further reference, we introduce the inverse lifetime, or escape rate, $r(C)$ of the state $C$, $r(C)=\sum_{C'\neq C} w(C\to C')$. 

As a consequence of these dynamics, a current flows from left to right through the system.
The current at time $t$, $j(i,t)$, gives the number of particles passing per unit of time through the bond between sites $i$ and $i+1$.
In a particular realisation, or {\it history}, of the TASEP, we can count the total integrated current $J(t,L)=\sum_{i=0}^L \int_0^t j(i,t')dt'\geq 0$ through all bonds up to time $t$ (where the particles entering respectively leaving the system correspond with $i=0$ and $i=L$). This is a stochastic variable whose properties can be obtained from the cumulant generating function
\begin{eqnarray}
\mu(s,t,L) = \frac{1}{t} \ln \langle e^{-s J(t,L)}\rangle
\label{2}
\end{eqnarray}
where the average $\langle \cdot \rangle$ is taken over the histories of the process. From (\ref{2}), the average current per unit time $\overline{J(t,L)}$ \footnote{overlined quantities represent averages over time, quantities with a $^\star$ represent time averages in the NESS.}, its variance $\overline{\Delta J (t,L)}$ and higher cumulants can be found by taking derivatives:
\begin{eqnarray}
\overline{J(t,L)} &=& \frac {1}{t} \langle J(t,L) \rangle =-\frac{\partial \mu}{\partial s} (0,t,L) \label{3}\\
\overline{\Delta J(t,L)} &=& \frac{1}{t} \left[ \langle J^2(t,L)\rangle - \langle J(t,L)\rangle^2\right] = \frac{\partial^2 \mu}{\partial s^2} (0,t,L)
\label{4}
\end{eqnarray}
Equation (\ref{2}) has a large mathematical similarity with the definition of the dimensionless free energy in equilibrium situations in which the inverse temperature $\beta=1/k_B T$ is replaced by $s$ and the sum over microstates is replaced by a sum over histories. 
We can therefore interpret (\ref{2}) as a {\it thermodynamics of histories} \cite{Lecomte07} in which realisations of the stochastic process are weighted by $\exp{(-sJ(t,L))}$. For negative $s$, (\ref{2}) will be dominated by histories with a large current, whereas for positive $s$ it will mainly get contributions from those with a small current.  There is no clear physical meaning to the variable $s$, but in the context of glassy dynamics, it has been shown that extending the parameter space with the variable $s$ can lead to new and interesting insights \cite{Merolle05,Jack06,Garrahan07}. 

In the long time limit, the TASEP evolves to a unique non-equilibrium steady state (NESS). The average current $J^*(L)=\lim_{t \to \infty} \overline{J(t,L)}$ in this state can be obtained exactly with a matrix technique \cite{Derrida93}. If one also takes the thermodynamic limit, the current per bond $J^\star=\lim_{L \to \infty} J^\star(L)/(L+1)$ is found to behave non-analytically as a function of $\alpha$ and $\beta$ leading to the recognition of three phases \cite{Derrida93}. In the low-density (LD) phase ($\alpha<1/2,\beta>\alpha$), $J^\star=\alpha(1-\alpha)$, while in the high-density (HD) phase ($\beta<1/2,\alpha>\beta$), $J^\star$ equals $\beta(1-\beta)$. Finally, in the maximal current (MC) phase ($\alpha>1/2,\beta>1/2$), $J^\star=1/4$. The TASEP therefore has three (boundary driven) nonequilibrium phase transitions.

It is useful to also introduce the $s$-weighted average current in the NESS together with its variance which are defined as
\begin{eqnarray}
J^\star(s,L) &=& - \lim_{t \to \infty} \frac{\partial \mu}{\partial s} (s,t,L) \label{3a}\\
\Delta^\star J(s,L) &=& \lim_{tÊ\to \infty} \frac{\partial^2 \mu}{\partial s^2} (s,t,L)
\label{4a}
\end{eqnarray}

For clarity, we now repeat a standard argument \cite{Lecomte07} that shows that in the NESS, the cumulant generating function equals the largest eigenvalue of a matrix $H(s)$, and that the approach to the asymptotic value is determined by the lowest gap of that matrix. We therefore introduce firstly the probability $P(C,J,t)$ that the system is in configuration $C$ and has integrated current $J$ at time $t$. Using (\ref{1}), we immediately find that
\begin{eqnarray}
\partial_t P(C,J,t)=\sum_{C'\neq C}\left[w(C'\to C) P(C',J-1,t)-w(C\to C') P(C,J,t)\right] 
\label{5}
\end{eqnarray}
Consequently, the discrete Laplace transform $\hat{P}(C,s,t) = \sum_{J=0}^\infty e^{-sJ} P(C,J,t)$ evolves according to
\begin{eqnarray}
\partial_t \hat{P}(C,s,t)=\sum_{C'\neq C}\left[w(C'\to C) e^{-s} \hat{P}(C',s,t) - w(C\to C')\hat{P}(C,s,t)\right]
\label{6}
\end{eqnarray}
To continue, it is convenient to introduce a matrix notation as is common in the so called 'quantum' approach to stochastic particle systems \cite{Schutz00}. We therefore introduce a set of basis vectors $|C\rangle$ each corresponding to a configuration $C$ and a vector $|\hat{P}(s,t)\rangle$ with components $\hat{P}(C,s,t)=\langle C|\hat{P}(s,t)\rangle$. Using this notation, the set of equations (\ref{6}) can be rewritten as
\begin{eqnarray}
\partial_t |\hat{P}(s,t)\rangle= H(s)|\hat{P}(s,t)\rangle
\label{7}
\end{eqnarray}
The diagonal elements of the matrix $H(s)$ are equal to minus the inverse lifetimes of the states, while the off-diagonal elements are given by the transition rates multiplied by $e^{-s}$. For $s=0$, (\ref{7}) reduces to the master equation (\ref{1}) and $H(0)$ corresponds to the generator of the stochastic process. We will therefore refer to $H(s)$ as the generalized generator.

Using the 'quantum' notation of \cite{Schutz00} in which an empty (occupied) site is associated with an up (down) spin, one can easily show that for the TASEP
\begin{eqnarray}
H(s) = \sum_{i=1}^{L-1} \left[ e^{-s} s_i^+s_{i+1}^ - - n_iv_{i+1}\right] + \alpha\left[e^{-s}s_1^- - v_1\right] + \beta \left[e^{-s} s_L^+ - n_L\right]
\label{8}
\end{eqnarray}
Here $n_i, v_i, s_i^+$ and $s_i^-$ are standard particle number, vacancy number, particle annihilation and creation operators at site $i$
\begin{eqnarray}
n=\left( \begin{array}{cc} 0\ \ 0 \\ 0\ \ 1 \end{array}\right),\ \
v=\left( \begin{array}{cc} 1\ \ 0 \\ 0\ \ 0 \end{array}\right),\ \
s^+=\left( \begin{array}{cc} 0\ \ 1 \\ 0\ \ 0 \end{array}\right),\ \
s^-=\left( \begin{array}{cc} 0\ \ 0 \\ 1\ \ 0 \end{array}\right)
\label{9}
\end{eqnarray}
The formal solution to (\ref{7}) is 
\begin{eqnarray}
|\hat{P}(s,t)\rangle = e^{H(s)t} |\hat{P}(s,0)\rangle
\label{10}
\end{eqnarray}
Therefore, we have
\begin{eqnarray}
\langle e^{-sJ(t,L)}\rangle &=& \sum_C \sum_J e^{-sJ(t,L)} P(C,J,t) \nonumber \\
&=&\sum_C \hat{P}(C,s,t)Ê\nonumber \\
&=&\sum_C \langle C|e^{H(s)t}|\hat{P}(s,0)\rangle
\label{11}
\end{eqnarray}
Using the spectral theorem, the sum in (\ref{11}) can be written in terms of the eigenvalues and eigenvectors of $H(s)$. In the long time limit, this sum is dominated by the largest eigenvalue $\lambda_1(s,L)$ while the first correction term involves the gap $G(s,L)=\lambda_1(s,L)-\lambda_2(s,L)$ with the second largest eigenvalue $\lambda_2(s,L)$.  One has
\begin{eqnarray}  
\langle e^{-sJ(t,L)}\rangle = A_1 e^{\lambda_1(s,L)t} \left[1 + A_2 e^{-G(s,L)t} + \cdots\right]
\label{12}
\end{eqnarray}
where $A_1$ and $A_2$ are time-independent factors depending on the initial conditions and the eigenvectors associated with the two largest eigenvalues. Therefore, for $tÊ\to \infty$
\begin{eqnarray}
\mu^\star(s,L)\equiv\lim_{t \to \infty} \mu(s,t,L)=\lim_{tÊ\to \infty} \frac{1}{t} \ln \langle e^{-sJ(t,L)}\rangle = \lambda_1(s,L)
\label{13}
\end{eqnarray}
which shows that $\mu^\star$ is intensive in time.

\section{Space-time phase transition}
In this section, we give a simple argument that shows that in the NESS and in the thermodynamic limit the cumulant generating function is non-analytic at $s=0$. Such phase transitions in the properties of histories have been called space-time transitions. 

Firstly, we observe that $\mu(s,t,L)$ is a non-increasing function of $s$ and by definition is zero at $s=0$. Hence we obtain the bounds
\begin{eqnarray}
0 \geq \mu(s,t,L) \geq \mu(s\to \infty,t,L) \hspace{2cm} s\geq 0
\label{16}
\end{eqnarray}
For $s\to \infty$, only histories with $J(t,L)=0$ contribute to $\mu$. This implies that the system at time $t$ is still in the configuration $C_0$ in which it was initially. The probability for this to happen decays exponentially with waiting time $1/r(C_0)$. Taking an average over possible initial conditions, one obtains 
\begin{eqnarray}
\lim_{s \to \infty} \langle e^{-sJ(t,L)}\rangle = \sum_{C_0} p_0(C_0) P(C_0,J=0,t) = \sum_{C_0} p_0(C_0) e^{-r(C_0) t}
\label{17}
\end{eqnarray}
where $p_0(C_0)$ is the probability that the system is at $t=0$ in $C_0$.
In the long time limit, this sum will be determined by the configurations with the largest lifetime.  It is not too difficult to realise that these are the completely empty configuration $C_e$ with $r(C_e)=\alpha$ and the fully occupied configuration $C_f$ with $r(C_f)=\beta$. Therefore, one has for very large times
\begin{eqnarray}
\lim_{s \to \infty} \langle e^{-sJ(t,L)}\rangle=  p_{min} e^{-\min[\alpha,\beta]\ t}\left[1 +  \frac{p_{max}}{p_{min}} e^{-|\alpha-\beta|t}+ \cdots\right]
\label{18}
\end{eqnarray}
(where $p_{min}=p_0(C_e)$ and $p_{max}=p_0(C_f)$ if $\alpha<\beta$, and $p_{max}=p_0(C_e)$ and $p_{min}=p_0(C_f)$ if $\alpha>\beta$.)
Comparing with (\ref{12}) we have that $\lambda_1(s\to\infty,L)=-\min[\alpha,\beta]$ and $G(s\to\infty,L)=|\alpha-\beta|$. Inserting (\ref{18}) in  (\ref{16}) and taking $t \to \infty$, one obtains
\begin{eqnarray}
0 \geq \mu^\star(s,L) \geq  -\min[\alpha,\beta]
\label{19}
\end{eqnarray}
These inequalities imply that the space-intensive quantities $\lim_{L\to \infty} \mu^\star(s,L)/(L+1)$ and $\lim_{L\to \infty}J^\star(s,L)/(L+1)$ are equal to zero for any strictly positive $s$. On the other hand, as discussed above, it is known from the exact solution \cite{Derrida93} that $J^\star=\lim_{L\to \infty} J^\star(0,L)/(L+1)$ is non-zero. Hence, it follows that in the $s$-ensemble, the TASEP has a first order space-time phase transition at $s=0$, and this for every $\alpha$ and $\beta$.

It is interesting to remark that this transition may be absent in the partially asymmetric exclusion process. In that model, particles can also jump to the left with a rate $q$ and enter the system on the right and leave at the left side. Hence no lower bound on $J(t,L)$ can be given and the argument presented above does not hold. 

In the remainder of this paper, we are interested in the finite size scaling properties of the cumulant generating function near the transition at $s=0$. For the TASEP on a ring it has been shown exactly \cite{Derrida98,Derrida99} that the cumulant generating function in the NESS scales as
\begin{eqnarray}
\mu^\star_{ring} (s,L) = sL \rho (1-\rho) + \sqrt{\frac{\rho(1-\rho)}{2\pi L^3}} H(s\sqrt{2\pi \rho (1-\rho) L^3})
\label{19b}
\end{eqnarray}
Here $\rho$ is the density of particles and $H$ is a scaling function. The first term on the right hand side of this equation equals $sLJ^\star$ since in the ring case $J^\star=\rho(1-\rho)$. In the remainder of this paper, we will use a numerical approach to investigate the scaling properties of $\mu$  for the case of open boundaries.

\section{The DMRG approach}
The behaviour of the cumulant generating function of the current at large times in a finite system is determined by the two largest eigenvalues of the generalized generator (\ref{8}). If we interpret this generator as a 'Hamiltonian' \cite{Schutz00}, we realise that calculating these eigenvalues is mathematically similar to determining the ground state energy and the gap of a quantum spin chain. The main difference from a standard quantum problem comes from the non-hermiticity of the generalised generator.

One of the most precise numerical procedures for calculating ground state properties of quantum chains is the density matrix renormalization group (DMRG) introduced by S White \cite{White92,White93,Schollwock05}. Later, this technique was adapted to generators of stochastic processes \cite{Hieida98,Carlon99} and in this context a study of the gap in the generator (at $s=0$) of the TASEP was made \cite{Nagy02}. More recently, we applied the DMRG for the first time to the generalized generators associated with current fluctuations in driven lattice gases and activity fluctuations in the contact process \cite{Gorissen09,Hooyberghs10}. 

The application of the DMRG to these generalized generators is not fundamentally different from the standard approach used for quantum systems. For some of the technical aspects, we refer to \cite{Hooyberghs10}.

We have used the DMRG to calculate the two largest eigenvalues of the generator (\ref{8}) as a function of $s$ for various points in the phase diagram of the TASEP. We are typically able to reach reliable results for $L$ up to $60$. At $s=0$ it is possible to go up to $L \approx 100$. The upper limit for $L$ that can be reached is essentially set by the stability of the Arnoldi algorithm used to diagonalise large non-Hermitian matrices.

Once the largest eigenvalue have been calculated with sufficient numerical accuracy, the average current and its variance in the $s$-ensemble are determined using numerical differentiation. Because of numerical round-offs, it is not possible to obtain cumulants beyond the second with sufficient accuracy.
\section{Current fluctuations in the stationary state}
In this section, we present our results for the current fluctuations in the NESS. 

\begin{figure}[t]
\begin{center}
\includegraphics[width=12.0cm]{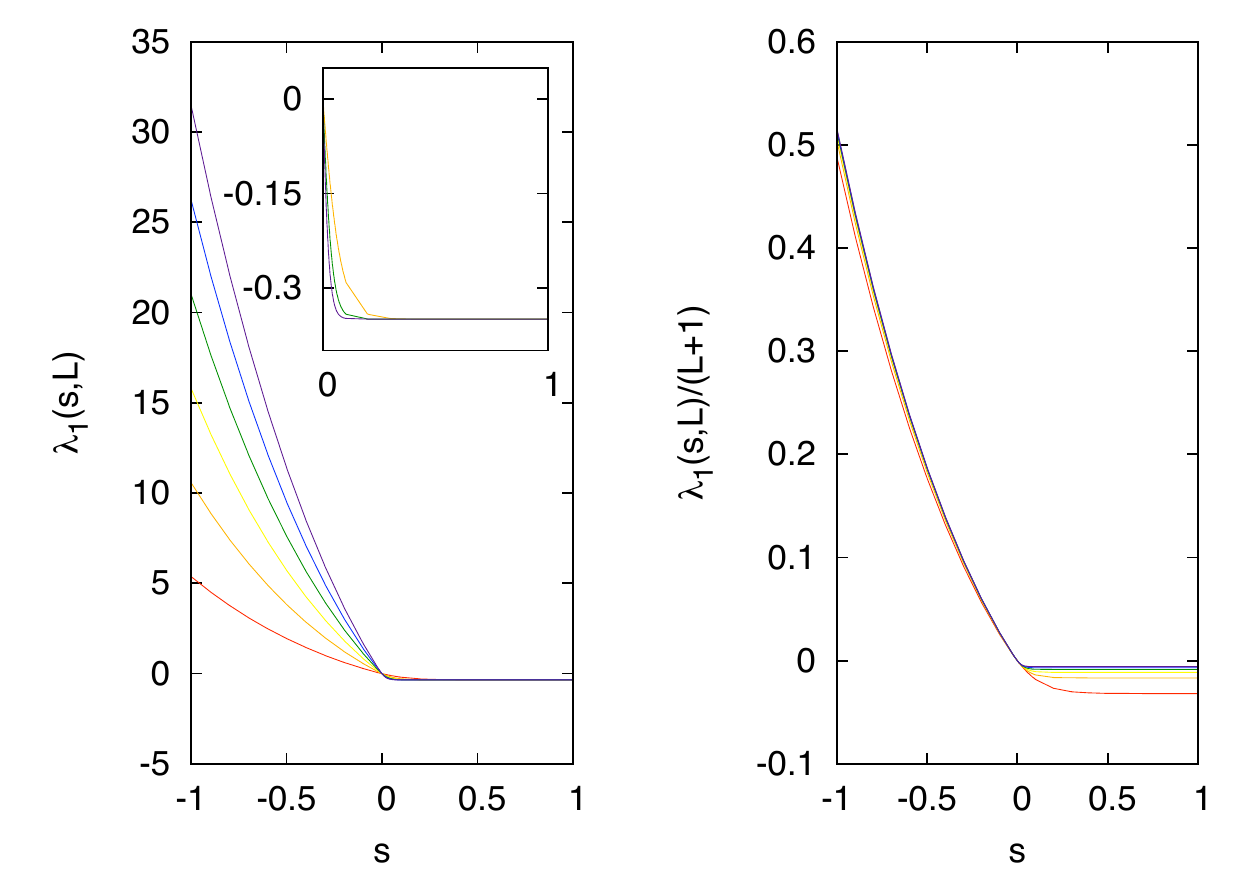}
\caption{\label{fig1} (Color online) Cumulant generating function as a function of $s$ for $\alpha=0.35, \beta=2/3$ for $L=10, 20, 30, 40, 50$ and $60$ (increasing from bottom to top for $s<0$ ). The region with $0\leq s\leq 1$ is shown in more detail in the inset ($L=20,40$ and $60$, increasing from top to bottom). On the right we show the CGF per bond for the same system sizes.}
\end{center}
\end{figure}

In Figure \ref{fig1} we show a typical result for the CGF $\mu^\star(s,L)=\lambda_1(s,L)$. The data are for a point in the low density phase with $\alpha=0.35,Ê\beta=2/3$. In the inset, it can be seen   that already for rather small $s$-values the asymptotic value $-\min[\alpha,\beta]=-0.35$ is reached, as predicted in the previous section. When $L$ increases this limiting value is reached for ever smaller $s$. The behavior shown here is typical for all values of $\alpha$ and $\beta$ that we investigated.

\begin{figure}[t]
\begin{center}
\includegraphics[width=12.0cm]{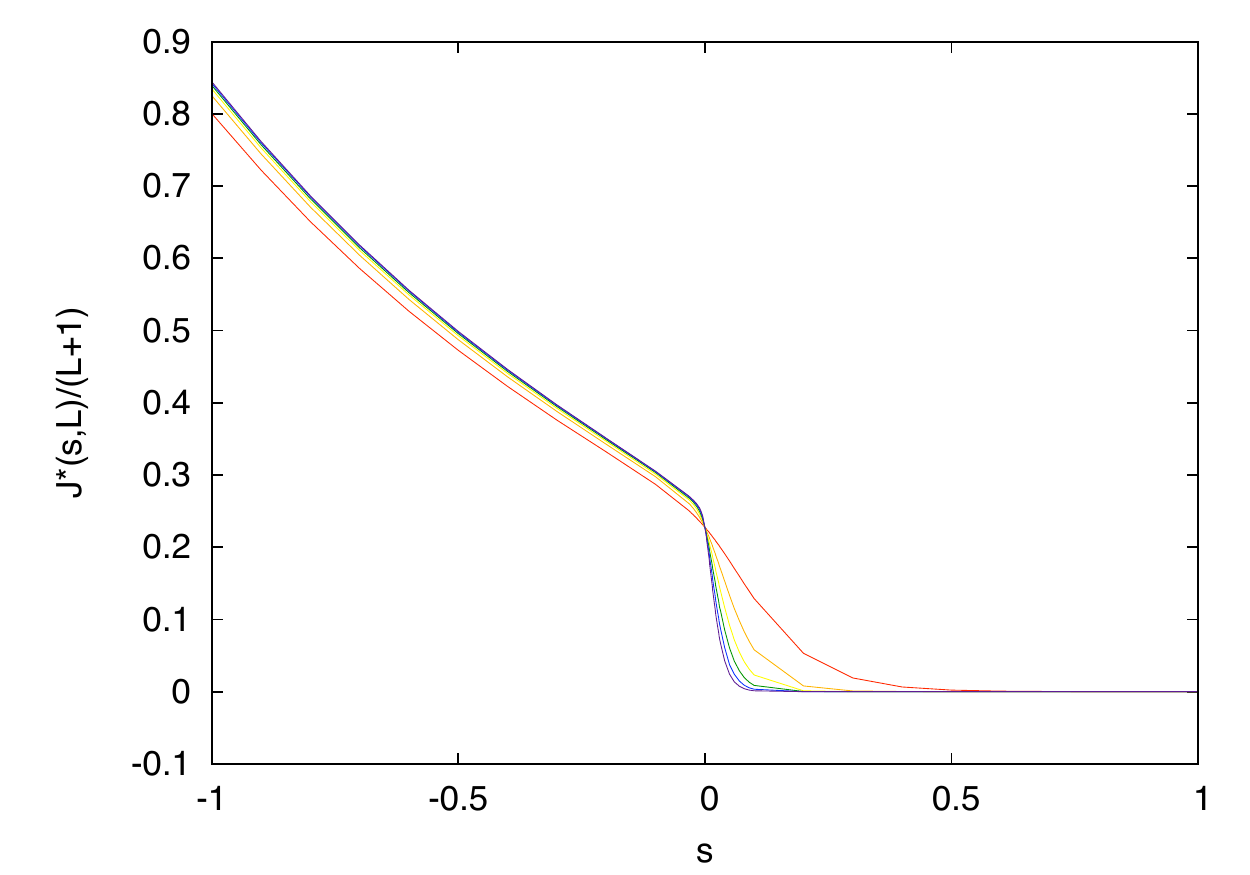}
\end{center}
\caption{\label{fig2} (Color online) Average current  per bond as a function of $s$ at $\alpha=0.50, \beta=2/3$ for $L=10, 20, 30, 40, 50$ and $60$ (increasing (decreasing) from bottom to top for $s$ negative (positive)).}
\end{figure}

In Figure \ref{fig2} we show the result for the average current per bond $J^\star(s,L)/(L+1)$ as a function of $s$ at $\alpha=.5, \beta=2/3$ (LD-MC transition line). In this figure, the space-time transition at $s=0$, rounded by finite size effects, is clearly visible. Qualitatively similar behaviour is found for other values of $\alpha$ and $\beta$.

We next propose a scaling form for the cumulant generating function. That such a scaling exists can be expected from analogy with the ring case, equation (\ref{19b}). Moreover, the similarity between the CGF and the equilibrium free energy leads us to expect that such a scaling can hold near the nonequilibrium phase transitions in the TASEP.  In writing down a scaling relation for the CGF a natural variable will therefore be $\Delta \alpha=\alpha - 1/2$, the distance from the transition between the LD and the MC phase (or equivalently $\Delta \beta=\beta-1/2$ for the HD to MC transition) \cite{Remark2}. These considerations lead us to make the finite size scaling ansatz
\begin{eqnarray}
\mu^\star(s,L,\Delta \alpha) =s(L+1)J^\star + L^{-z} H(sL^{y_s},\Delta \alpha L^{y_\alpha})
\label{20}
\end{eqnarray}
where $H$ is a scaling function and $y_s$ and $y_\alpha$ are two critical exponents that we will determine below. The factor $L+1$ in the first term on the right hand side equals the number of bonds through which the total current passes (including incoming and outgoing particles). It replaces the factor $L$ in the ring case. The appearance of the dynamical exponent $z$ in the prefactor $L^{-z}$  is a consequence of the time intensivity of $\mu^\star$. It replaces the space-dimension $d$ in the factor $L^{-d}$ appearing in the scaling of the equilibrium free energy.

From (\ref{20}) it follows that the average current at $s=0$ scales as
\begin{eqnarray}
J^\star (s=0,L,\Delta \alpha) = (L+1)J^\star + L^{-z+y_s} H'(\Delta \alpha L^{y_\alpha})
\label{21}
\end{eqnarray}
where $H'$ is another scaling function.
We can compare this scaling prediction with the exact results for large $L$.
Firstly, at the LD-MC transition one has \cite{Derrida93} 
\begin{eqnarray}
J^\star (s=0,L,\Delta \alpha=0)= \frac{L+1}{4}( 1 + \frac{1}{2L} + \cdots )
\label{22}
\end{eqnarray}
Comparing with (\ref{21}), one obtains $y_s=z=3/2$. To also determine $y_\alpha$ we take the derivative of (\ref{21}) with respect to $\Delta \alpha$ at $\alpha=1/2$. We get
\begin{eqnarray}
\frac{d J^\star}{d\alpha}(s=0,L,\Delta \alpha=0) \sim L^{y_\alpha} 
\label{23}
\end{eqnarray}
From the exact asymptotic results in the LD-phase \cite{Derrida93}, one can easily find that asymptotically in $L$
\begin{eqnarray}
\frac{d J^\star}{d\alpha}(s=0,L,\Delta \alpha=0)=\frac{\sqrt{\pi}}{2} L^{1/2}(1+\cdots)
\end{eqnarray}
so that we conclude $y_\alpha=1/2$.
Inserting the exponent values into (\ref{21}), we obtain the final form for the scaling of the CGF near the LD-MC transition.
\begin{eqnarray}
\mu^\star(s,L,\Delta \alpha) =s(L+1)J^\star + L^{-3/2} H(sL^{3/2},\Delta \alpha L^{1/2})
\label{24}
\end{eqnarray}

We have performed several tests of this ansatz with the DMRG-approach and where possible using exact results. 

\begin{figure}[t]
\begin{center}
\includegraphics[width=12.0cm]{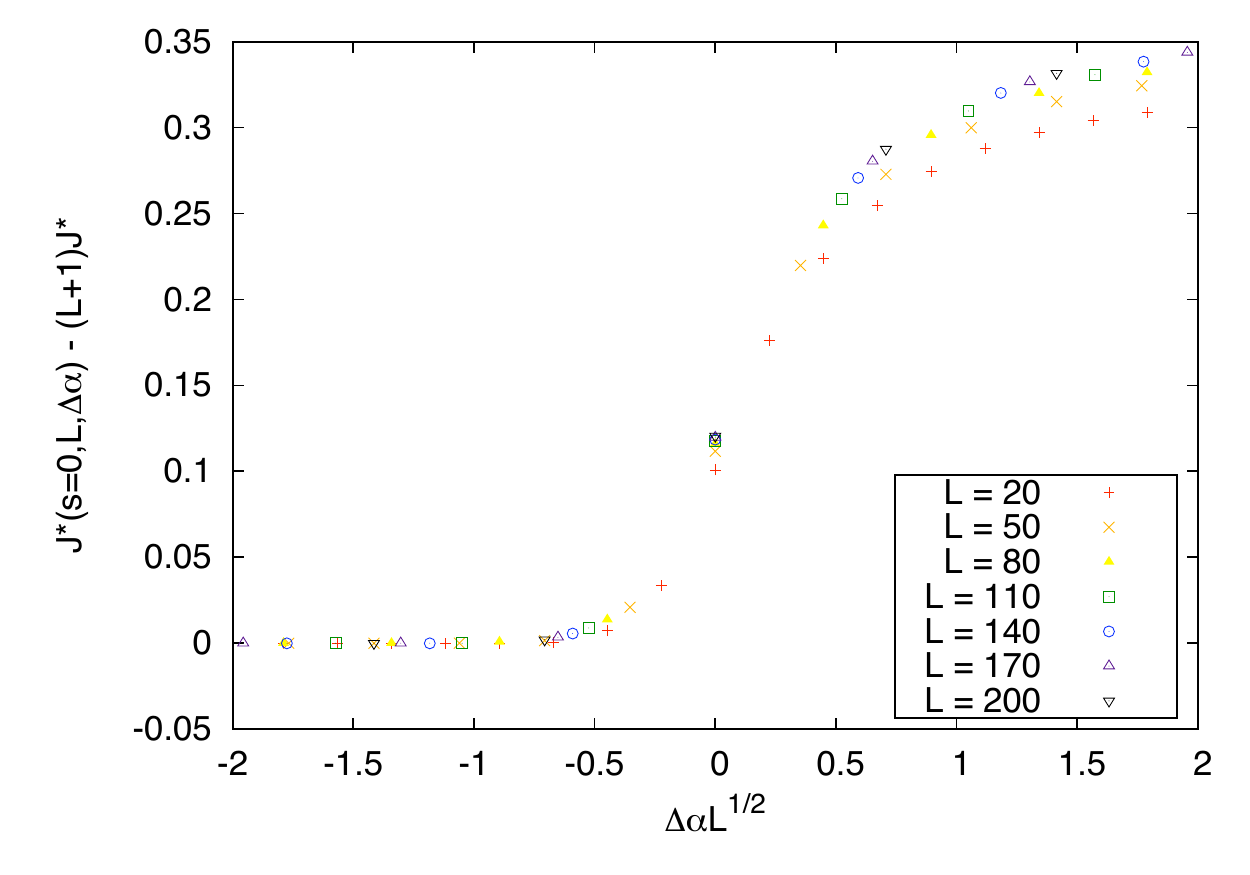}
\end{center}
\caption{\label{fig3bis} (Color online) Scaling of $J^\star(s=0,L,\Delta \alpha) - (L+1)J^\star$ as a function of $\Delta \alpha L^{1/2}$ at $\beta=2/3$.}
\end{figure}

We start with the scaling for the current at $s=0$ as given in (\ref{21}) since it can be checked using the results in \cite{Derrida93}. In Figure \ref{fig3bis} we present data obtained by numerical evaluation of the exact expression for the current for $L$ up to 200. As can be seen the scaling is well satisfied, especially from $L \approx 80$ on.  

\begin{figure}[t]
\begin{center}
\includegraphics[width=12.0cm]{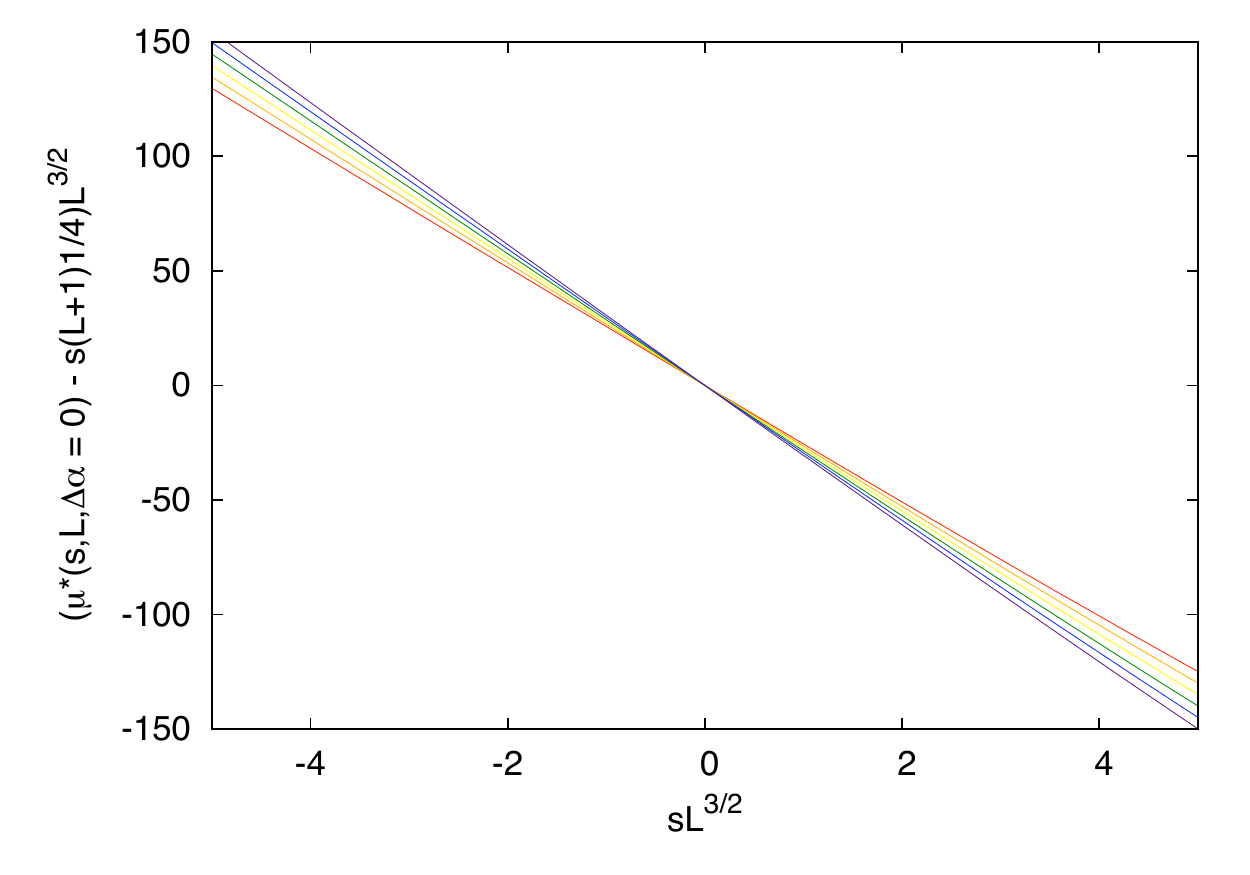}
\end{center}
\caption{\label{fig4} (Color online) Scaling of $(\mu^\star(s,L,0) - s(L+1)/4)L^{3/2}$ as a function of $sL^{3/2}$ at $\alpha=1/2, \beta=2/3$ (for $L=50, 52, 54, 56, 58$ and $60$ (increasing (decreasing) from top to bottom for positive (negative) $s$).}
\end{figure}

Next we turn to the scaling of the CGF itself. Figure \ref{fig4}  shows a scaling plot of $(\mu^\star(s,L,0) - s(L+1)/4)L^{3/2}$ as a function of $x=sL^{3/2}$. These data are determined from the DMRG. We used the data for small $s$ and the largest $L$ values that we could obtain ($L=60$). The curves for various $L$-values are seen to come closer together with increasing $L$. 
Unfortunately, the $L$-values that can be studied with the DMRG at $s \neq 0$ seem to be just outside the scaling region.

\begin{figure}[t]
\begin{center}
\includegraphics[width=12.0cm]{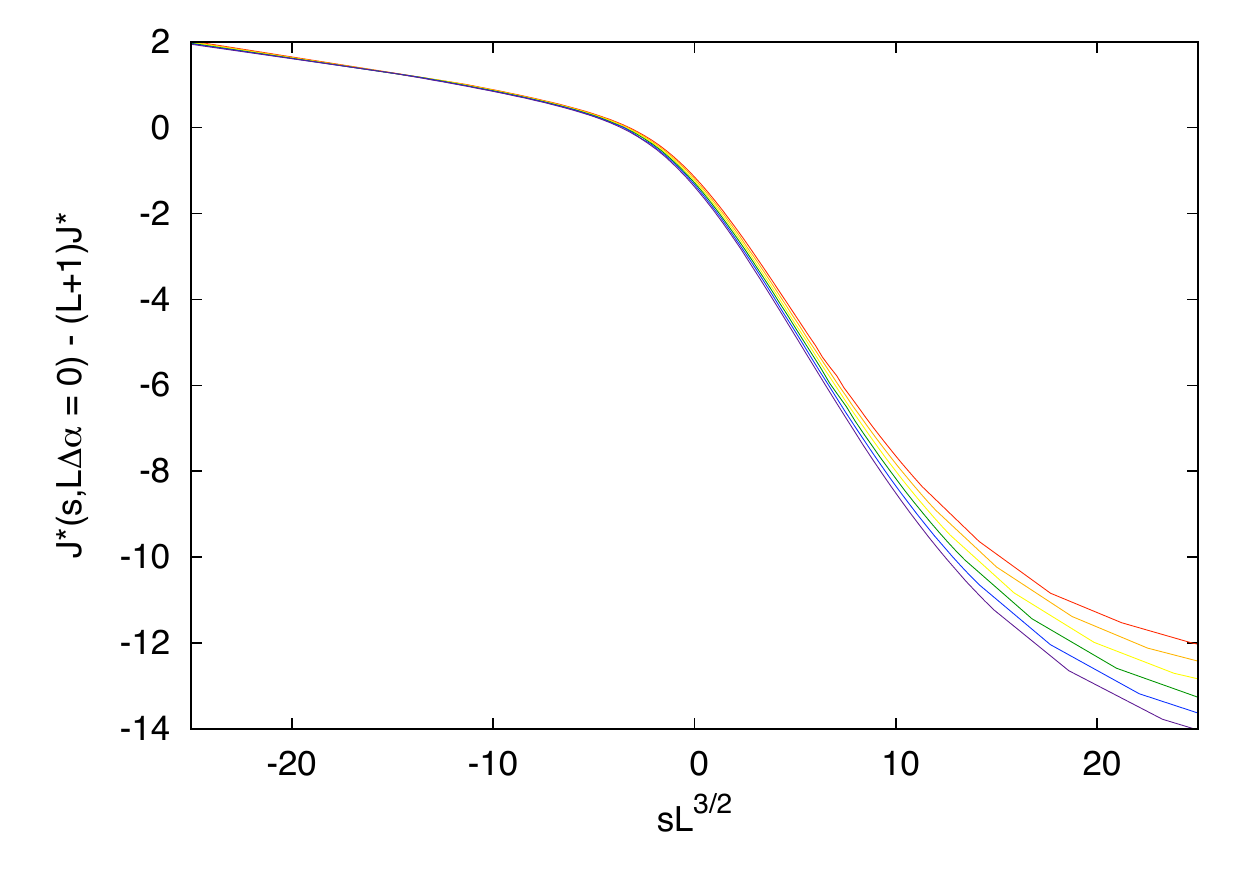}
\end{center}
\caption{\label{fig3tris} (Color online) Scaling of $J^\star(s,L,\Delta \alpha=0) - (L+1)J^\star$ as a function of $s L^{3/2}$ (for $L=50, 52, 54, 56, 58$ and $60$ increasing from top to bottom).}
\end{figure}

In Figure \ref{fig3tris} we present the scaling plot for $J^\star(s,L)$ as a function of $s$ at $\Delta\alpha=0$ from the DMRG data. There is a good collapse of the data, especially at negative $s$-values.

\begin{figure}[t]
\begin{center}
\includegraphics[width=12.0cm]{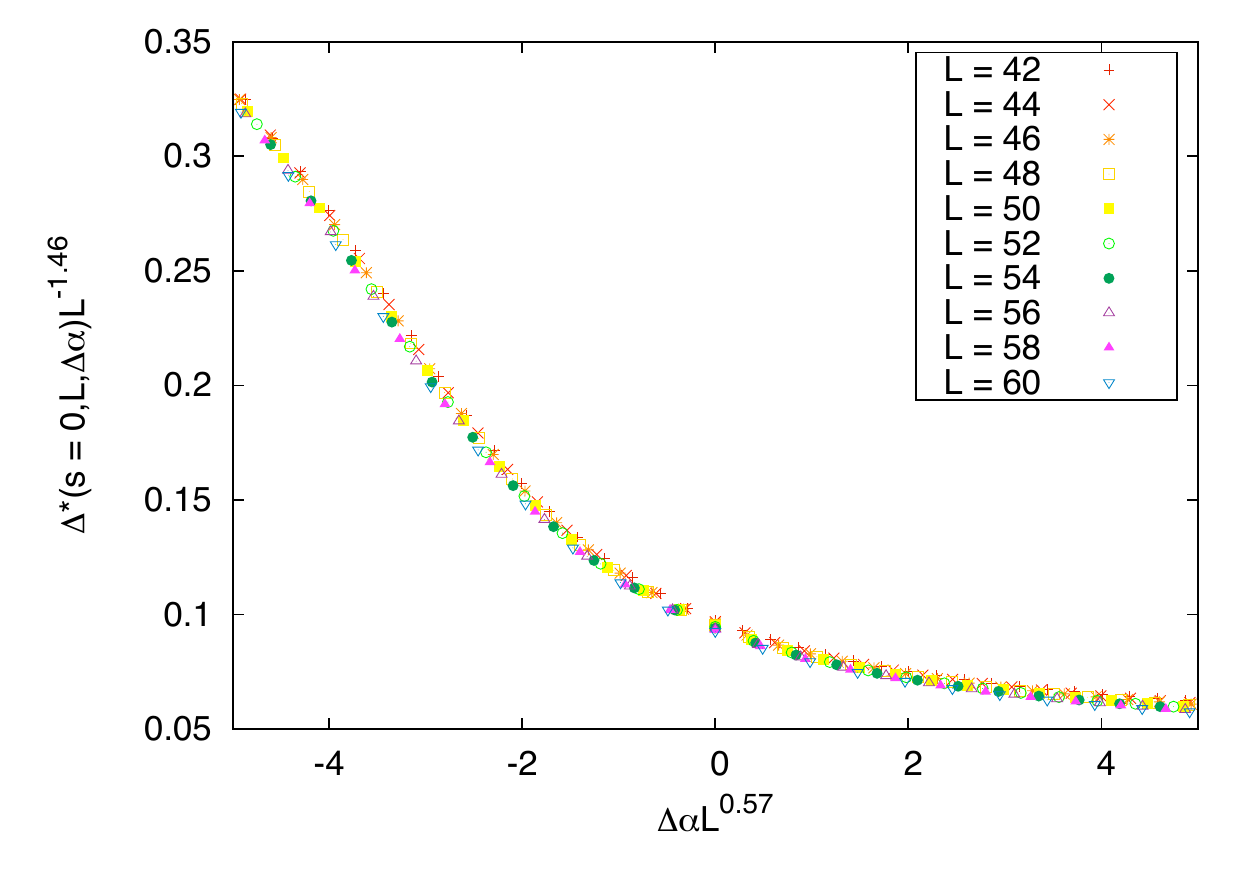}
\end{center}
\caption{\label{fig5} (Color online) Scaling of  the variance of the current at $\beta=2/3$. The numerical data have been collapsed with the algoritm of \cite{Somendra01}. The best collapse shown here is found for $y_\alpha=.57$ and $z-2y_s=1.46$.}
\end{figure}

Finally, we turn to the variance of the current, which in this non-equilibrium context plays a role similar to the specific heat or the susceptibility in equilibrium. How does the variance of the current behave near the LD-MC transition? From (\ref{24}), we derive the scaling form at $s=0$
\begin{eqnarray}
\Delta^\star J(0,L,\Delta \alpha)= L^{3/2} H_2(\Delta \alpha L^{1/2})
\label{27}
\end{eqnarray}
with $H_2$ a scaling function.
An exact formula for the variance of the current was derived in \cite{Derrida95} using the matrix product technique. Due to the complexity of this formula, closed expressions for the variance could only be derived at the point $\alpha=\beta=1$ in the MC-phase, and along the line $\alpha+\beta=1$ (LD and HD phase). For these situations it was found that $\Delta^\star J$ grows respectively as $L^{3/2}$ and as $L^2$. We calculated the variance of the current in several points in the phase diagram using the DMRG technique and verified that these results are universal for each phase. Moreover, at the LD-MC transition line we found that $\Delta^\star J(0,L,\Delta \alpha) \sim L^{1.50\pm.02}$, consistent with (\ref{27}) \cite{Gorissen09}. To be in agreement with the scalings in the LD and MC phase just discussed, $H_2(x)$ should be constant for $x \gg 0$ and $H_2(x) \sim x$ for $x \ll 0$. In Figure \ref{fig5} we show our results for $\Delta^\star J(0,L,\Delta \alpha)L^{z-2y_s}$ as a function of $\Delta \alpha L^{y_\alpha}$. Here we used the algorithm presented in \cite{Somendra01} to collapse the data for the largest $L$-values. As the figure shows, this is possible but  with exponents that deviate ten percent from the conjectured ones. This is again because our data are not yet in the asymptotic regime where scaling is expected to hold. The prediction of linear behaviour for  $H_2(x)$ when $x$ is negative enough is however clearly verified.

From (\ref{24}) one can also obtain a scaling form for the variance of the current as a function of $s$. We have checked that also this form is consistent with the DMRG-results as already shown in Figure 3 of \cite{Gorissen09}.

In conclusion, the scaling assumption (\ref{24}) allows a consistent description of all exact and numerical data on the average current and its fluctuations. Due to the symmetries of the TASEP we expect a completely similar scaling to hold near the HD-MC transition.

\section{Time dependent current fluctuations}
We first discuss our results for the gap $G(s,L)$ near the LD-MC transition and then investigate their implications for the time-dependent current flucutations.

The behaviour of the gap for $s=0$ was first determined numerically using the DMRG approach to stochastic operators \cite{Nagy02}. Later exact results were obtained from an analysis of the Bethe ansatz equations \cite{deGier05,deGier06}. In the MC phase, it was found that $G(0,L)$ goes to zero as $L^{-3/2}$, consistent with the criticality of the phase. In the LD and HD phase on the other hand, the gap goes to a constant indicating a finite relaxation time. Here we study for the first time the gap as a function of $s$. 

\begin{figure}[t]
\begin{center}
\includegraphics[width=12.0cm]{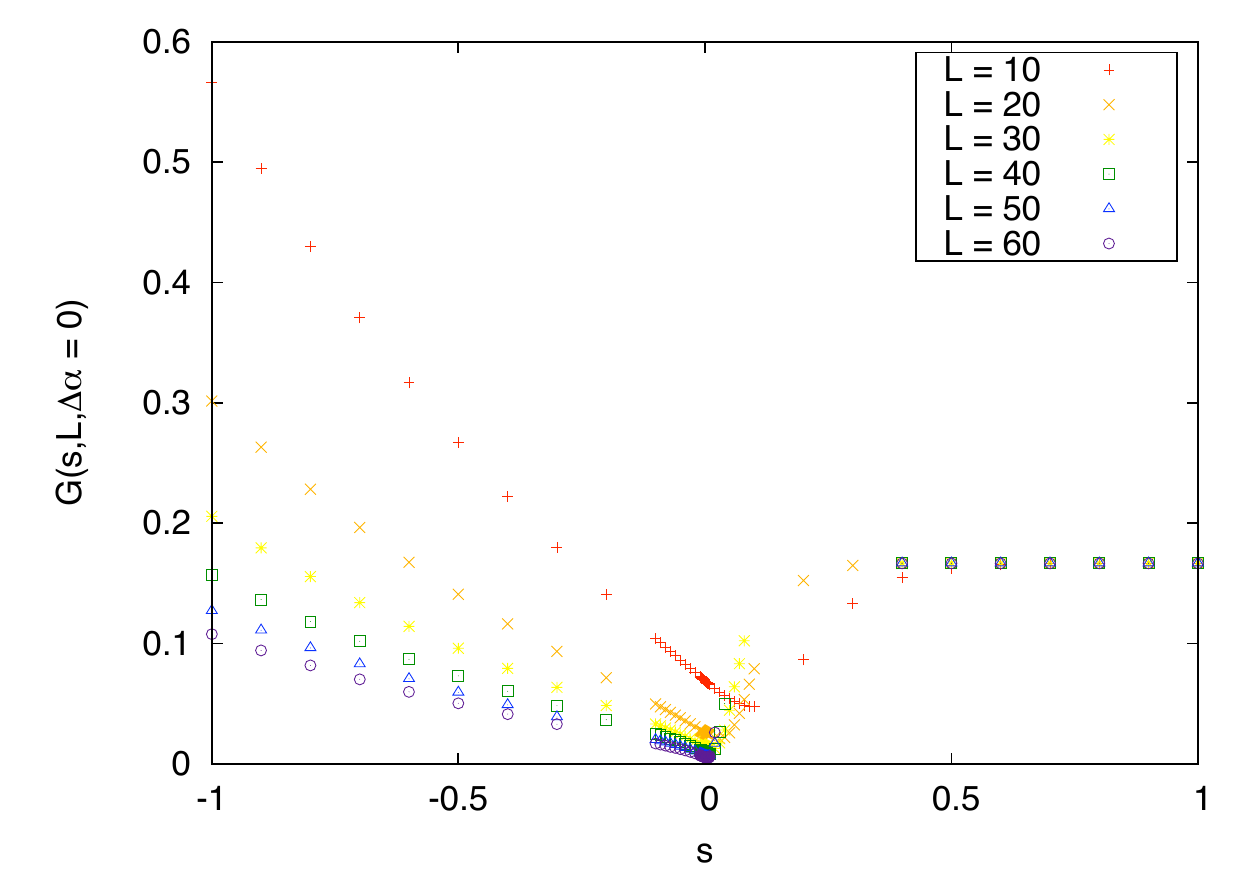}
\end{center}
\caption{\label{fig6} (Color online) Gap $G(s,L)$ as a function of $s$ for different $L$-values and for $\alpha=1/2, \beta=2/3$.}
\end{figure}

In Figure \ref{fig6}, we show as a typical result for different $L$-values at $\alpha=1/2, \beta=2/3$. As can be seen the gap approaches $|\alpha-\beta|=1/6$, the value predicted  for large $s$ in section 3, already for rather small $s$-values. 

\begin{figure}[t]
\begin{center}
\includegraphics[width=12.0cm]{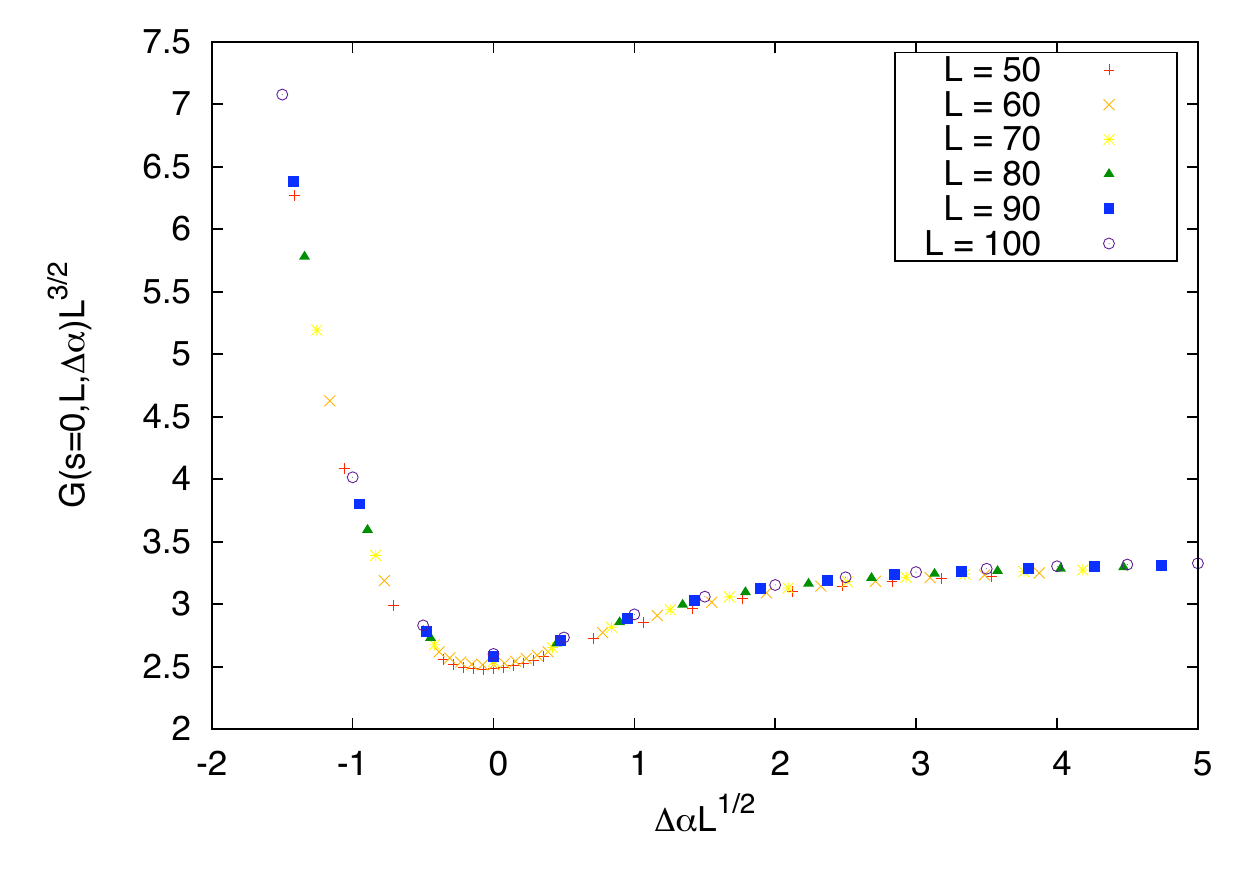}
\end{center}
\caption{\label{fig7} (Color online) Scaled gap $G(0,\Delta\alpha,L)L^{3/2}$ as a function of $\Delta\alpha L^{1/2}$.}
\end{figure}

We have found that all the exact and numerical results for the gap can again be described by a scaling function. In analogy with (\ref{24}) we propose the form
\begin{eqnarray}
G(s,L,\Delta \alpha) = L^{-3/2} F(sL^{3/2},\Delta\alpha L^{1/2})
\label{27bis}
\end{eqnarray}
where $F$ is a scaling function. This form can describe the exact results at $s=0$ if $F(0,y)$ goes to a constant for $y\gg0$ and goes as $y^3$ for $y\ll0$. Figure \ref{fig7} shows our numerical results. At $s=0$ reliable values for the gap can be obtained up to $L \approx100$. The scaling is already satisfied very well for $L \approx 50$ and the data also show the expected behaviour for large $|y|$.

\begin{figure}[t]
\begin{center}
\includegraphics[width=12.0cm]{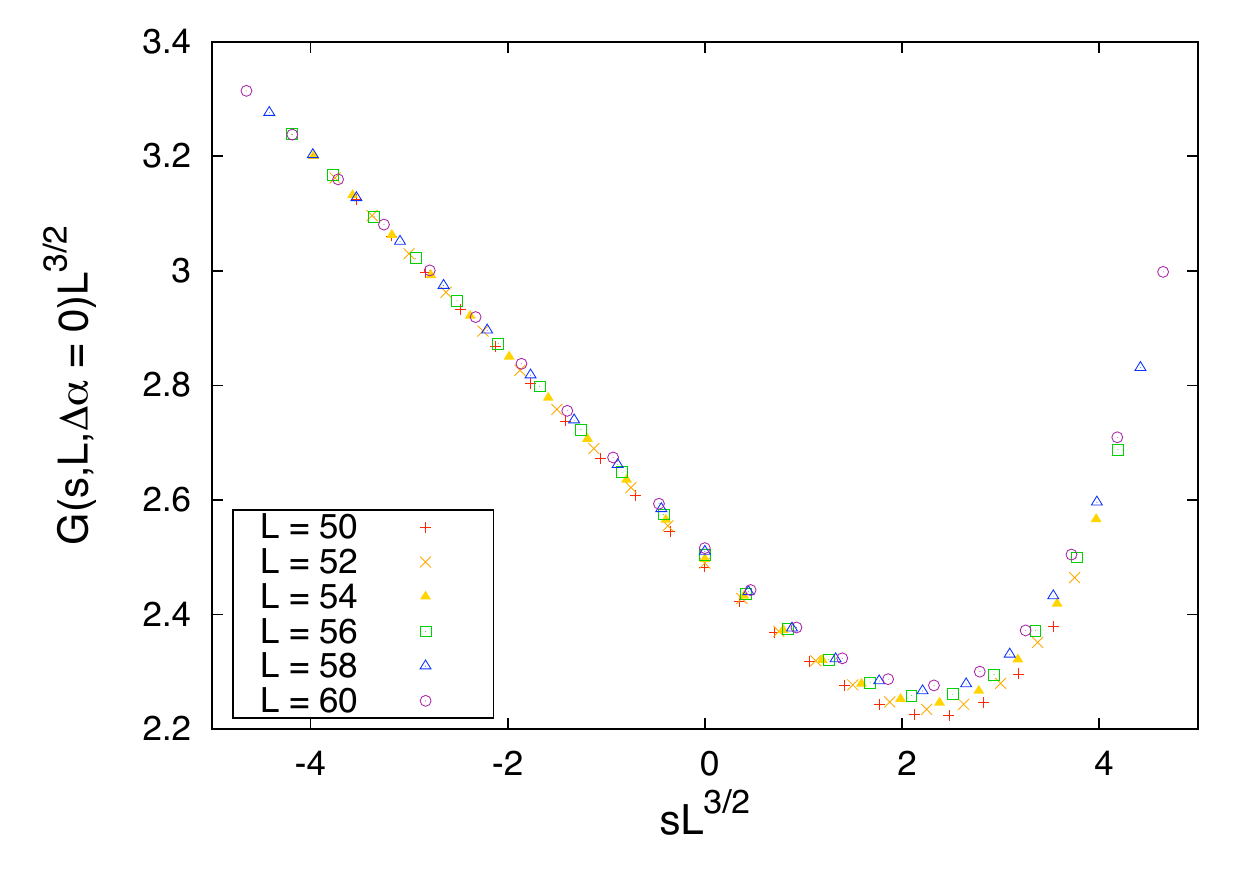}
\end{center}
\caption{\label{fig8} (Color online) Scaled gap $G(s,L) L^{3/2}$ at the LD-MC transition as a function of $sL^{3/2}$. }
\end{figure}

In Figure \ref{fig8}, we show the scaling of the gap at $\Delta \alpha=0$ as a function of $s$. There is again a good collapse, especially for negative $s$-values.

\begin{figure}[t]
\begin{center}
\includegraphics[width=12.0cm]{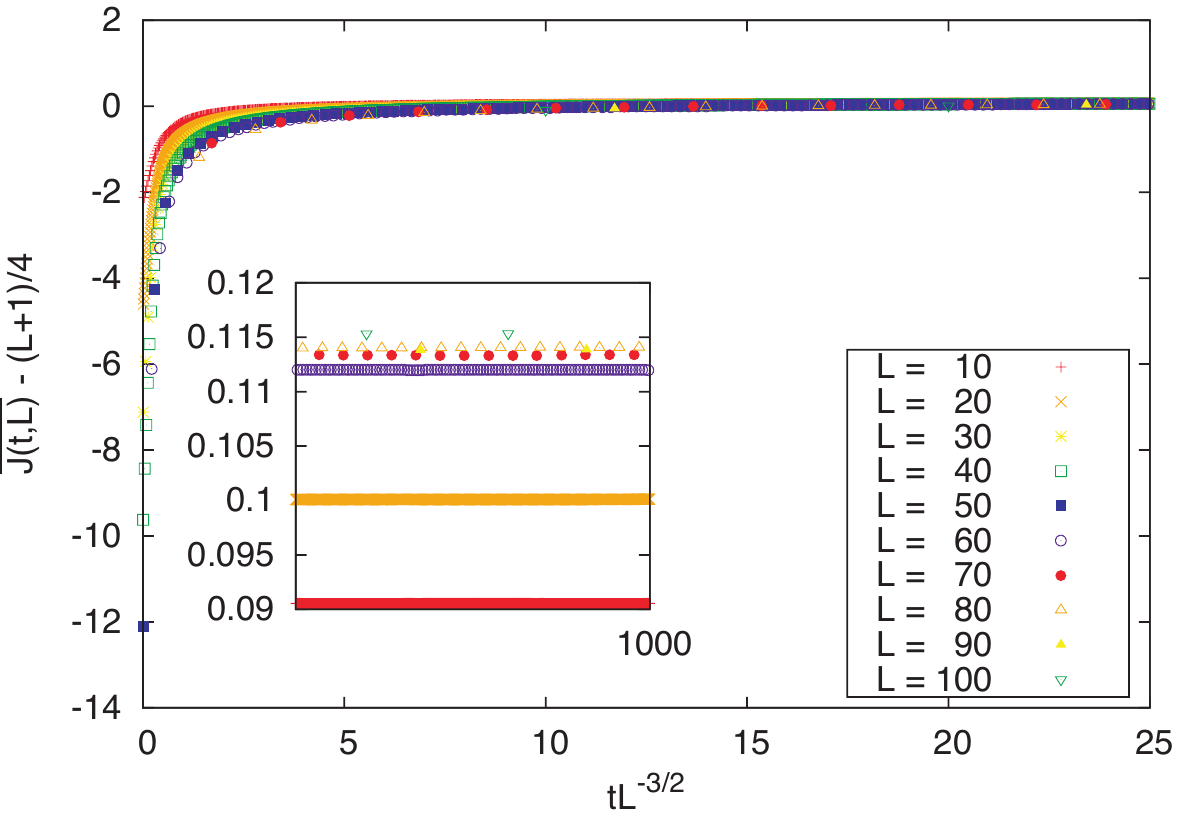}
\end{center}
\caption{\label{fig9} (Color online) Scaling plot of $\overline{J(t,L,\Delta\alpha=0)} - (L+1)/4$ as a function of $tL^{-3/2}$. The main figure shows the result for intermediate values of the scaling variable. The inset shows the data for large $t$.}
\end{figure}

The scalings that we have found for $\mu^\star$ and $G$ imply a scaling for the late time behaviour of the time-dependent current fluctuations. Inserting (\ref{24}) and (\ref{27bis}) into (\ref{12}), we obtain
\begin{eqnarray*}
\hspace{-1.6cm}\langle e^{-sJ(t)}\rangle=A_1 e^{s(L+1)J^\star t} \left[ e^{L^{-3/2}t H(sL^{3/2},\Delta\alpha L^{1/2})}\left(1+A_2 e^{-L^{-3/2}tF(sL^{3/2},\Delta\alpha L^{1/2})}+\cdots\right)\right]
\label{28}
\end{eqnarray*}
The whole term within square brackets depends on time through the combination $tL^{-3/2}$, so that we obtain
\begin{eqnarray}
\mu(s,L,t)=s(L+1)J^\star + L^{-3/2} K(sL^{3/2},\Delta\alpha L^{1/2},tL^{-3/2})
\label{29}
\end{eqnarray}
where $K$ is once more a scaling function. This equation gives insight in how the time-dependent current and its fluctuations reach their asymptotic NESS value. For the average current at the LD-MC transition, we get for example
\begin{eqnarray}
\overline{J(t,L,\Delta\alpha=0)} = (L+1)/4 + K_1(tL^{-3/2})
\label{30}
\end{eqnarray}
where $K_1$ is another scaling function. Comparison with (\ref{22}) gives $K_1(x) \to 1/8$ for $x \to \infty$. For the variance of the current one finds
\begin{eqnarray}
\overline{\Delta J(t,L,\Delta \alpha)} = L^{3/2} K_2(\Delta \alpha L^{1/2}, tL^{-3/2})
\label{31}
\end{eqnarray}
The time dependent current and its variance cannot be obtained directly from our DMRG-method. Here instead, we have verified relations like (\ref{30}) and (\ref{31}) by simulations based on the Gillespie algorithm for $L$ up to 100. In Figure (\ref{fig9}) we show our data for $\overline{J(t,L,\Delta\alpha=0)} - (L+1)/4$ as a function of $tL^{-3/2}$. The initial condition was an empty lattice. We simulated $10^4$ histories for $L \leq 50$. For larger system sizes the number of realisations decreased up to $2000$ for $L=100$. As can be seen the scaling is well satisfied. At large times, the current reaches its $L$-dependent NESS value, which for $L \to \infty$ should reach the asymptotic value $1/8$ (inset).

\begin{figure}[t]
\begin{center}
\includegraphics[width=12.0cm]{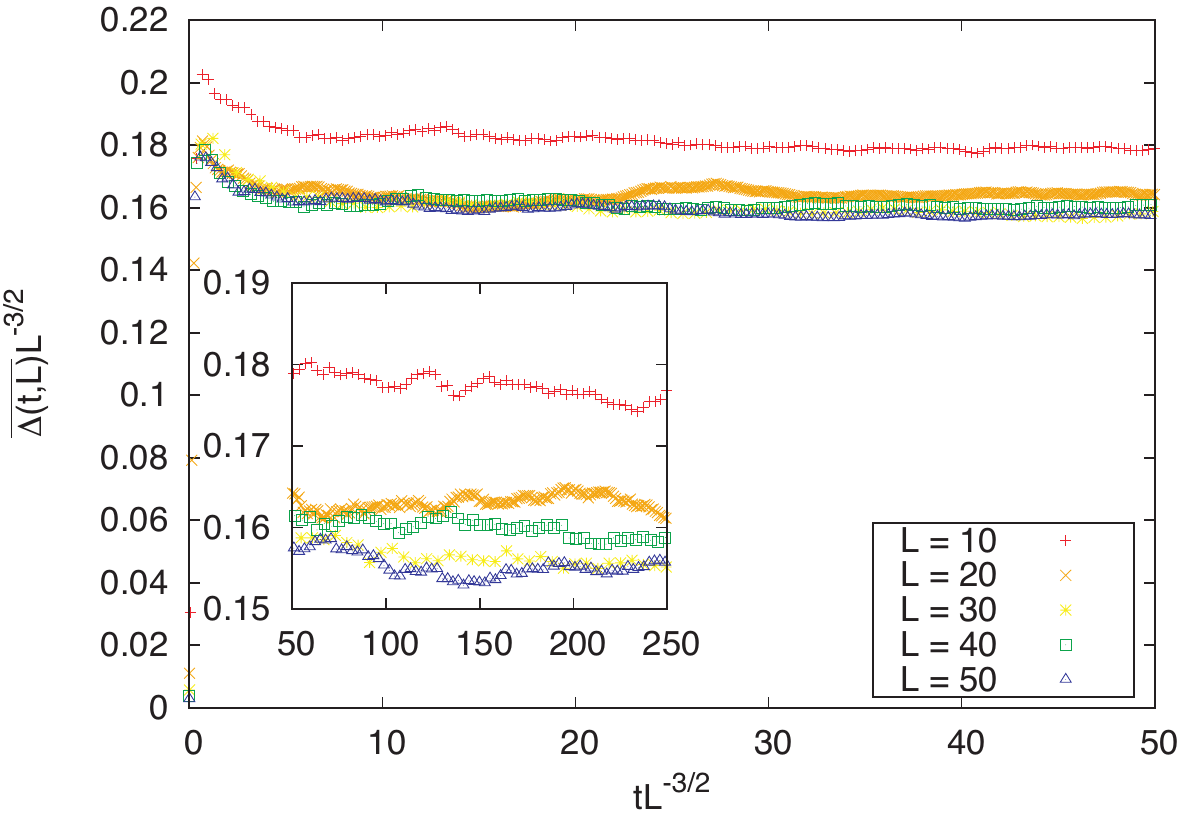}
\end{center}
\caption{\label{fig10} (Color online) Scaling plot of $\overline{\Delta J(t,L,\Delta\alpha=0)}L^{-3/2}$ as a function of $tL^{-3/2}$ at $\alpha=1/2,\beta=2/3$. The main figure shows the result for intermediate values of the scaling variable. The inset shows the data for $t$ large.}
\end{figure}

A similar picture arises for the time-dependence of the variance of the current though the data are somewhat noisier  for the number of histories that we could simulate (see Fig. \ref{fig10}).
Within their accuracy they are consistent with the prediction (\ref{31}). The variance of the current in the NESS cannot be obtained very precisely from these simulations. While consistent with the DMRG-values, the latter are much more precise. This is another advantage of the DMRG in comparison with simulations.
\section{Conclusions}
In this paper we have studied the current fluctuations of the TASEP with open boundaries, both in the NESS and as a function of time. 

We have shown that, in the NESS and in the thermodynamic limit, the cumulant generating function must be non-analytical at $s=0$ implying the existence of a space-time phase transition in the TASEP.

We have proposed the finite size scaling form (\ref{29}) for the (time-dependent) cumulant generating function. Important variables herein are the distance $\Delta \alpha$ from the non-equilibrium phase transition line and $s$. The form proposed is a natural extension of that near equilibrium phase transitions and that found for the TASEP on the ring. 

We have verified our ansatz using available exact results and numerical approaches based on the DMRG and the Gillespie algorithm. 

A scaling function for the current generating function has also been derived recently for the symmetric simple exclusion process (SSEP) on a ring \cite{AppertRolland08}. The form for the CGF derived in that work shows some similarity with our ansatz (\ref{20}) when, as should be expected for a diffusive system, the dynamical exponent $z$ is changed to 2. It would be of interest to investigate how general our scaling form is.

The TASEP was originally introduced as a model for mRNA-translation \cite{MacDonald69}. We have recently shown how fluctuations in the number of proteins produced by one mRNA can be related to the time-dependent current fluctuations in a model for translation \cite{Gorissen10}. From the scaling proposed here, it is then possible to derive how, for example, the variance in the number of proteins produced depends on mRNA-length. This opens a road to a possible experimental verification of our scaling ansatz.

In this paper we have only studied the two largest eigenvalues of the generalised generator $H(s)$. The DMRG also allows the calculation of the associated left- and right eigenvectors. From these it is possible to obtain expectation values, such as the density at a given site \cite{Garrahan09}. By tuning the parameter $s$, it is then possible to see how the typical density profile changes for a current that deviates from its average value. We are currently calculating these properties. The results will be published elsewhere.\\
\ \\
{\bf Acknowledgement} We thank V. Lecomte for useful discussions.


\begin{thebibliography}{10}
\bibitem{Remark2}
We couldn't study the so called coexistence line, which is the first order transition line between the LD and HD phase, because of numerical instabilities in the DMRG procedure.
\end{thebibliography}
\end{document}